# Bending of chiral fractal lattice metamaterials

Wenjiao Zhang[a,b*], Robin Neville[b], Dayi Zhang[c,b], Jie Yuan[d], Fabrizio Scarpa[b*], Roderic Lakes[e]

[a]School of Engineering, Northeast Agriculture University, No. 600 Changjiang Road, Harbin, 150030, China

[b]Bristol Composites Institute, University of Bristol, Bristol BS8 1TR, UK

[c]School of Energy and Power Engineering, Beijing University of Aeronautics and Astronautics (BUAA), Beijing, 100191, China

[d]Mechanical and Aerospace Engineering, University of Strathclyde, James Weir Building, 75 Montrose Street, Glasgow G1 1XJ, UK

[e]Department of Engineering Physics, University of Wisconsin-Madison, 1500 Engineering Drive, Madison, WI 53706-1687, USA

## Abstract

We describe the out-of-plane bending of chiral fractal lattices metamaterials by using a combination of theoretical models, full-scale finite elements and experimental tests representing the flexural behaviour of metamaterial beams under three-point bending. Good agreement is observed between the three sets of results. Parametric analyses show a linear log-log relation between bending modulus and aspect ratios of the unit cells, which are indicative of the fractal nature of the metamaterial. The ratio between the bending and in-plane tensile moduli of these chiral fractal metamaterials ranges between ~ 5 and ~ 34 and is linearly proportional to the square of the ratio between length and width of the ribs of the chiral unit cells at different fractal orders. These properties suggest that the class of chiral fractal lattice metamaterials offer metacompliance properties between the flexural and in-plane stretching behaviours, which can be tailored by the adoption of the fractal scales.

## 1. Introduction

---

[*] Corresponding Authors: Wenjiao Zhang (zhangwenjiao@neau.edu.cn) and Fabrizio Scarpa (f.scarpa@bristol.ac.uk)

Topologies related to mechanical metamaterials largely involve the use of lattices and/or multiphase composites, as well as applying patterns of perforations to convert conventional materials substrates into architected materials. Examples of classes of metamaterials systems include hexagonal lattices [1], rotating rigid units (rotating squares, rectangles, parallelograms, rhomboidal, triangles, cubes, and hierarchical mixed topologies) [2-13], chiral pattern formations [14], as well as re-entrant configurations [15]. Regular or random slits/cuts have also been adopted to generate unusual mechanical behaviour in mechanical metamaterials [16-19]. Many two-dimensional metamaterials have been studied. Relatively little work has however been performed on the out-plane behaviour of metamaterials. Examples involve the design and testing of a hybrid auxetic foam/perforated plate structure [20], in which the static bending stiffness of the hybrid auxetic composite is modelled using analytical and finite element (FE) approaches benchmarked against three-point bending experimental tests. Another example is the evaluation of the out-of-plane bending and energy dissipation of chiral perforated lattices beams under quasi-static and large deformations three-point (3P) bending [21].

Chiral cellular structures with rotational symmetry are a subset of auxetic (i.e., negative Poisson's ratio) solids. Chiral hexagonal configurations have been first made, measured, and analysed in the pioneering works of Lakes' group, in non-affine deformations [22] and later further developed into an in-plane chiral hexagonal topology [23]. The through-the-thickness properties of chiral structures are significant for the multifunctional behaviour of sandwich panels. Some of these mechanical characteristics involve the flexural stiffness, energy absorption, resistance to wrinkling as well as sound and electromagnetic insulation [24]. Ref [25-27] have described the elastic buckling behaviour of hexagonal and tetrahedral chiral honeycombs under flatwise compressive loading. Lorato et al [28] have investigated the out-of-plane linear elastic mechanical properties of trichiral, tetrachiral and hexachiral honeycomb configurations, and developed finite element models to identify the dependence of the transverse shear stiffness versus the gauge thickness of the honeycombs. Alderson et al [29] have evaluated the out-of-plane bending deformation of four types of honeycombs with trichiral, anti-trichiral, re-entrant trichiral as well as re-entrant anti-trichiral configurations by finite element simulations. Chen [30, 31] developed theoretical models for evaluating the flexural rigidity and torsional rigidity of an open honeycomb. The models involved the identification of a torsion coefficient for an equivalent thin plate in bending by using a generalized variational principle. Constraint conditions on the edges of a unit cell plate were considered, as a part of the analysis related to the torsional deformation occurring during the out-of-plane deformation of the honeycomb. Hou et al [32] have described the bending and failure of polymorphic honeycomb auxetic topologies consisting of gradient variations of the horizontal rib length and the cell internal angles across the surface of the cellular structures. Ha, Plesha and Lakes [33] have also developed and modelled chiral three-dimensional cubic lattices with rigid cubical nodules using finite elements. The 3D chiral lattices exhibit stretch-twist coupling, which cannot happen in a classical elastic continuum, but occur in a chiral Cosserat solid [34]. Huang et al [35] have presented a series of analytical models, finite element simulations and experimental tests to evaluate the bending performance of zero Poisson's ratio (ZPR) cellular structures made from the tessellation of hexagons and thin plates.

This paper describes the out-plane bending mechanical properties of a chiral fractal perforated metamaterial using theoretical models benchmarked with finite element results and experimental three-point bending tests. Chiral fractal perforated configurations are sometimes classified within metamaterials manufactured via kerfing techniques [36, 37], and have recently shown some remarkable bandgap tuning capabilities under extreme out-of-plane deformations [38]. The analytical and finite element models of the flexural stiffness in chiral fractal metamaterials presented in this work are related to equivalent thin plates/beam configurations, different from the deep beams described in Ref [21], which undergo internal deformation mechanisms involving visco-hyperelasticity of the core material and contact friction through the thickness, when subjected to large displacements. The generalised analytical model of the flexural modulus of the chiral fractal metamaterial presented in this work are benchmarked by an asymptotic homogenization model using full-scale finite element simulations of the three-point tests and related experimental results applied to metamaterial beams spanning two fractal orders. The effect of the geometric parameters of the chiral fractal unit cell on the effective bending modulus of the metamaterials is evaluated for four fractal orders by using theoretical computations and compared with finite elements techniques. We also show that the chiral fractal metamaterial configuration described in this work provides a bending to axial stiffness ratio from 6 to 35, depending upon the fractal order considered. The paper is organized as follows: a theoretical bending model of the unit cell of the chiral hinge lattice will be firstly presented, followed by the description of the setup of the experimental tests and the related full finite element simulations. After that, the effective bending modulus obtained from the theoretical, numerical and experimental data are compared in detail. The effects of the different aspect ratios of the chiral fractal metamaterial are also discussed.

## 2. Theoretical bending model of the chiral fractal metamaterial

Figure. 1 shows the square unit cell of the chiral fractal metamaterial with parameters *a*, *b*, *t* and *h*; the stress contour of the related finite element model is representative of the out-plane bending deformation of the metamaterials. The geometry of the metamaterial is based on the lattice configuration presented in [39]. The architecture of the chiral fractal metamaterial consists in self-similar generations of a series of cuts like the first iteration of a Peano's curve [40]. The in-plane mechanical properties of the metamaterial are dictated by the in-plane bending, stretching and transverse shear of the ribs. Within the geometry of this class of metamaterials, the parameter *a* represents the whole length of unit cell, *b* is the elementary width of the ribs, which is equal to 2mm in the test cases of this paper. The term *h* is the out-plane thickness of the unit cell and *t* is the width of the slit. The total length of the unit cell is $a=n\times(b+t)$, where *n* is an even integer (10 in cell represented in Fig. 1). The ratio *a/b* is critical to determine the homogenised in-plane Young's and shear moduli of the metamaterial, with in-plane stiffness decreasing by almost an order of magnitude when passing from *a/b*=6, to *a/b*=10 [39]. In the analytical bending model proposed, the value of the gap *t* is simplified as zero. The same gap *t* is however equal to *b*/10 in the finite element models used here and consistent with the cutting parameters adopted in the fabrication of the experimental samples made via laser cutting.

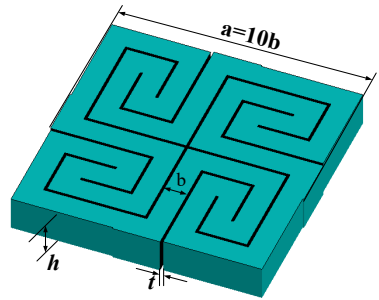

(a)

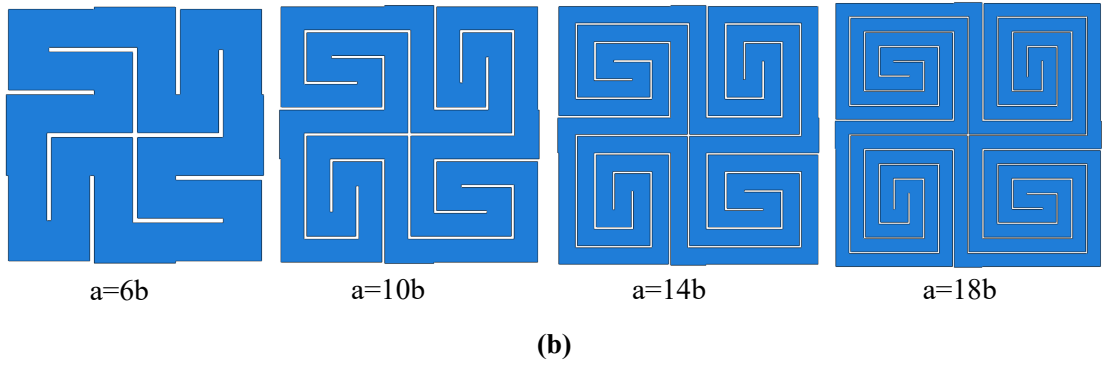

| a=6b | a=10b | a=14b | a=18b |

**(b)**

**Fig. 1. (a) Geometric parameters of a chiral fractal lattice unit cell (a). The four fractal configurations evaluated in this work (b).**

The simplifications and assumptions under out-plane loading are like those used in the previous work related to the in-plane mechanics of the chiral fractal metamaterial [39]. The single unit cell of the chiral fractal metamaterial is represented by a series of Timoshenko beams (36, in the case of the a/b=10 configuration shown in Fig. 1). For the out-plane bending loading, the boundary conditions are six degrees of freedom fixed on the left-hand corner [30, 35] and a uniform load distribution $q_z$ applied to the right area adjacent to the neighbour unit cell (Fig.2 (a)). The mechanical properties of the unit cell are therefore simplified by evaluating the deformation behaviour of an equivalent closed, sequential, and statically indeterminate beam structure with cross-section parameters $b$ and $h$, with the uniform loading $q_z$ replaced by an equivalent concentrated force $P_z$ (Fig. 2 (b)). The left corner between the beam components 1 and 36 is further cut open to solve for the internal forces $F_z$, $T$ and $M$ in the beam element 1, with complementary conditions $\delta_1^z = \delta_1^\varphi = \delta_1^\theta = 0$ (Fig. 2 (c)).

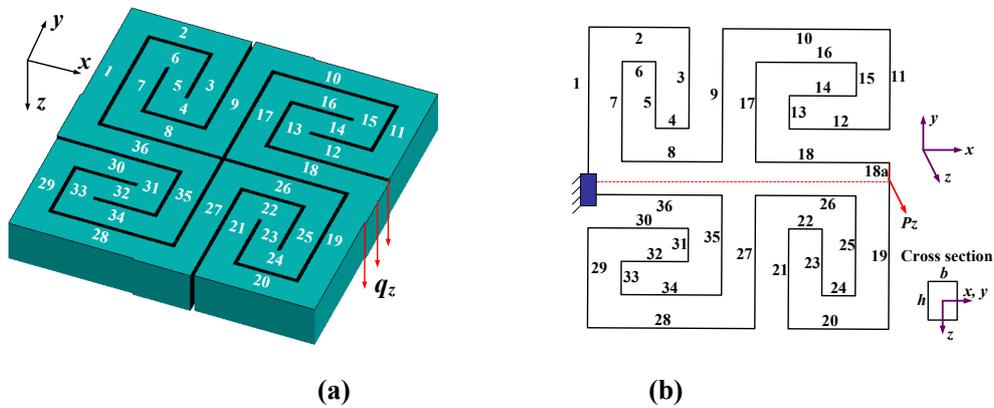

**(a)** **(b)**

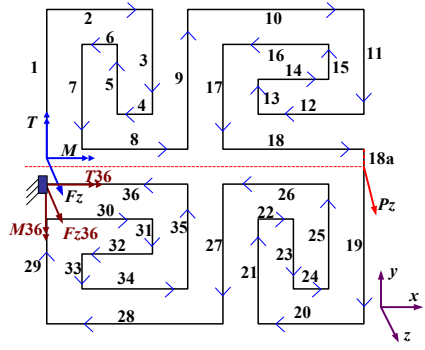

(c)

Fig. 2. Schematic illustration of the out-plane theoretical modelling of the unit cell a=10b: (a) element division; (b) structure simplification; (c) statically determinate structure with complementary conditions

The signs of the internal forces in each beam element under out-plane loading are defined in the following way. The positive direction for the shear force $F_{zi}$ is parallel to the cross-section area of the beam and along the $z$ axis. The positive rotation direction for bending moment $M_i$ and torque $T_i$ is along $x$, $y$ axis based on the right-hand rule, respectively. Similarly to the derivation of the in-plane internal force equations for the chiral lattice metamaterial [39], the calculation of internal forces firstly begins from beam element 1, as follows:

$$M_1(x_1) = F_z \times x_1 + M, \quad F_{s1}(x_1) = F_z, \quad T_1(x_1) = T, \quad 0 \leq x_1 \leq l_1 \qquad (1)$$

Then, the other beam elements can be represented by their previous elements correspondingly. For the beam element $i$ ($i = 2..18, 18a..36$), the torque $T_i(x_i)$ and shear force $F_{si}(x_i)$ can be represented by the out-plane bending moment $M_{i-1}(l_{i-1})$ and shearing force $F_{si-1}(l_{i-1})$ of beam element $i$-1:

$$\begin{cases} F_{si}(x_i) = F_{si-1} \\ F_{s19}(x_{19}) = F_{s18a} + P_z \end{cases}, \quad \begin{cases} T_i(x_i) = M_{i-1}(l_{i-1}) \\ T_{19}(x_{19}) = T_{18a} \end{cases}, \quad i = 2..18a, 20..36 \qquad (2)$$

The out-plane bending moment equations of the other beam elements can be expressed as $M_i(x_i) = T_{i-1}(l_{i-1}) \pm F_{Si-1} \times x_i$, $i = 2..36$. The sign here has following conventions: (1) when the beam $i$ is turning downwards or turning right towards the beam $i+1$, the sign in the $i^{th}$ bending equation is negative; (2) when the beam $i$ is directing upwards or turning left towards beam $i+1$, the sign in the $i^{th}$ equation is positive (Fig. 2 (c)). Consequently, the out-plane bending moment equations of the other 35+1 beam elements are rewritten as:

$$\begin{cases} M_i(x_i) = T_{i-1} - F_{si-1} \times x_i & \rightarrow or \downarrow \\ M_{19}(x_{19}) = M_{18a}(l_{18a}) - F_{s18a} \times x_{19} - P_z \times x_{19} \\ M_i(x_i) = T_{i-1} + F_{si-1} \times x_i & \leftarrow or \uparrow \end{cases} \quad (3)$$

The effective flexural modulus of the chiral fractal metamaterial structure is then calculated by applying Castigliano's second theorem [41]. In one unit cell, each element undergoes out-plane bending moment $M_i(x)$, torque $T_i(x)$ and shear loading $F_{si}(x)$. The total strain energy of one unit cell structure is therefore defined as:

$$U = \sum U_i = \sum_{i=1}^{n} \left( \int_0^{l_i} \frac{M_i^2(x)}{2E_c I} dx + \int_0^{l_i} \frac{T_i^2(x)}{2GI_P} dx + k \int_0^{l_i} \frac{F_{Si}^2(x)}{2GA} dx \right) \quad (4)$$

Where $E_c, G = E_c/(2(1+v))$ are the Young's and shear moduli of the core material and $v$ is Poisson's ratio of the same core solid. The parameter $k$ is the shear coefficient factor for a Timoshenko beam, defined as $k = \frac{10(1+v)}{12+11v}$ [42]. The cross-section area of each element is $A = hb$, while $I = bh^3/12$ is the second moment of area. The polar moment of inertia of a solid with a rectangular cross section is $I_P = \frac{hb^3}{16}\left(\frac{16}{3} - \frac{3.36b}{h}\left(1 - \frac{b^4}{12h^4}\right)\right)$, $h > b$ [41].

Substituting Eq. (1)-Eq. (3) to Eq. (4), the total strain energy of the chiral fractal lattice cell under a concentrated out-of-plane loading $P_z$ is then obtained as:

$$U = \frac{P_z^2}{E_c h} \times \frac{\sum_{i=0}^{8}\left(\alpha_i v^i \times \sum_{j=0}^{35} \beta_j \left(\frac{b}{h}\right)^j\right)}{\sum_{i=0}^{7}\left(\gamma_i v^i \times \sum_{j=0}^{33} \zeta_j \left(\frac{b}{h}\right)^j\right)} \quad (5)$$

Where $\alpha_i, \beta_j, \gamma_i$ and $\zeta_j$ are constants. From Eq. (5), the displacement $\delta_z$ under the out-plane bending loading $P_z$ is:

$$\delta_z = \frac{\partial U}{\partial P_z} = \frac{2P_z}{E_c h} \times \frac{\sum_{i=0}^{8}\left(\alpha_i v^i \times \sum_{j=0}^{35} \beta_j \left(\frac{b}{h}\right)^j\right)}{\sum_{i=0}^{7}\left(\gamma_i v^i \times \sum_{j=0}^{33} \eta_j \left(\frac{b}{h}\right)^j\right)} \quad (6)$$

The effective flexural modulus of a cantilever beam structure for unit cell can be therefore calculated as:

$$E_f = \frac{Pl^3}{3wI} = \frac{P_z(10b)^3}{3\delta_z \times (10bh^3/12)} = E_c \times \frac{\sum_{i=0}^{7}\left(c_i v^i \times \sum_{j=0}^{35} d_j \left(\frac{b}{h}\right)^j\right)}{\sum_{i=0}^{8}\left(e_i v^i \times \sum_{j=0}^{35} g_j \left(\frac{b}{h}\right)^j\right)} \quad (7)$$

Where $c_i$, $d_j$, $e_i$ and $g_j$ are constants. From Eq. (7), it is possible to evince that the non-dimensional effective bending modulus $E_f/E_c$ is dependent upon the ratio of width to out-plane thickness $b/h$ of the unit cell and the Poisson's ratio of the core material $v$ alone.

## 3. Manufacturing and experimental tests

All chiral fractal metamaterial samples used in this paper have been manufactured using a laser cutting facility applied to PMMA substrates (World Lasers LR1612 laser cutter with a 40W $CO_2$ laser). The elastic properties of the PMMA plastic have been determined by dog-bone specimens according to the standard test method (ASTM D638-08) in our previous work [39]. Two types of samples with unit cell *a*=6*b* and *a*=10*b* of 280mm×60mm×3mm were manufactured for three-point bending experiment. The samples had at least 12×3 cells along the main *x* and *y* directions (Fig. 3). The Young's modulus of the PMMA is $E_c$=2.23±0.26GPa, with a Poisson's ratio of *v*=0.37±0.02. Those data have been used in the theoretical and finite element simulations. According to the ASTM D790 standard test, the 280 mm long specimens allow for 50mm overhanging at each end, which is at least 10% of the support span and sufficient to prevent the specimen from slipping through the supports. A support span of 180mm is used for these specimens, which has a span-to-depth ratio of 60:1 to eliminate shear effects when the data are used to identify the equivalent flexural modulus. Three-point bending tests of the chiral fractal metamaterial samples have been performed using a Instron 3343 test machine with a 1 KN load cell and a constant displacement rate of 5mm/min. The tests were stopped when the central point deflection reached 10mm.

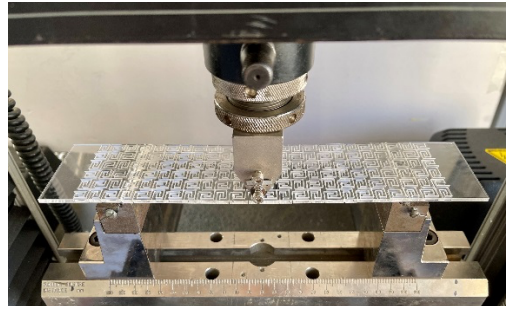

(a)

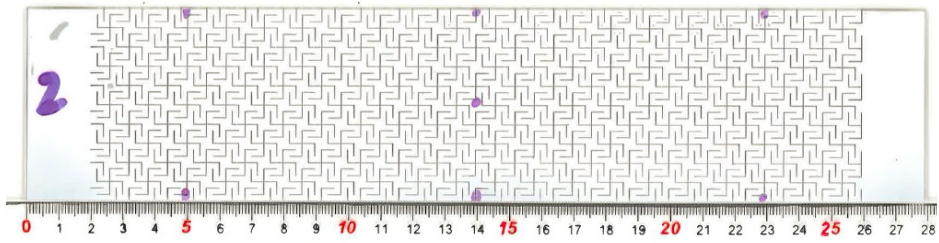

(b)

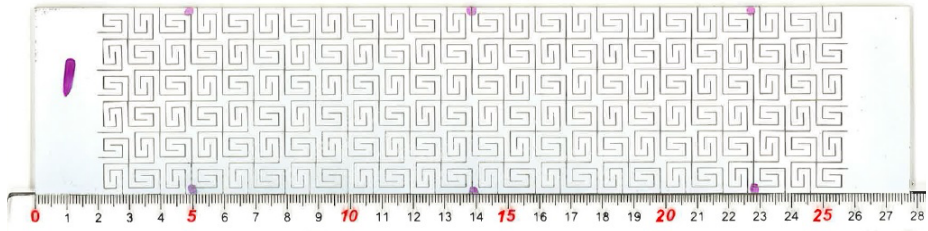

(c)

**Fig. 3. Experimental setup of the three-point bending tests (a) and two types of specimens used in the tests: (b) unit cell a=6b, (c) unit cell a=10b**

The tangent flexural modulus in the linear elastic regime is obtained as following:

$$E_f = \frac{L^3 m}{4wh^3} \qquad (8)$$

Where $E_f$ is the modulus of elasticity in bending, $L$ is the support span, $w$ is the width of beam tested, $h$ is the out-plane depth of tested beam, and $m$ is slope of the tangent to the initial straight-line portion of the load-deflection curve.

## 4. Finite Element simulations

The numerical simulation of the chiral fractal lattice under three-point bending tests have been performed using ABAQUS version 6.14 implicit. The full-size geometric dimensions and the material properties of the finite element models were the same as those of the experimental specimens described above. Three discrete rigid shells were placed on the top and bottom of the chiral fractal metamaterial plate, regarded as one indenter and two supporters (Fig. 4). The metamaterial plate was meshed with linear hex-dominated 8-node bricked solid elements (C38DR) and the three rigid shells were then meshed with a 4-node 3D bilinear rigid quadrilateral element (R3D4). An element size of $t/4=0.5$mm with sweep technique meshing was applied to guarantee the numerical convergence of the results. The simulations involved static with nonlinear geometric deformations options. A general contact model based on the Coulomb friction and defined as All* with self was applied, with a frictional penalty coefficient of 0.2 along the tangential directions, as well as a hard normal contact [43, 44]. The two lower support shells were constrained with fixed boundary conditions without motion and rotation, the upper indenter was applied with a constant loading displacement of 10mm. The equivalent bending modulus was then calculated by using Eq. (8).

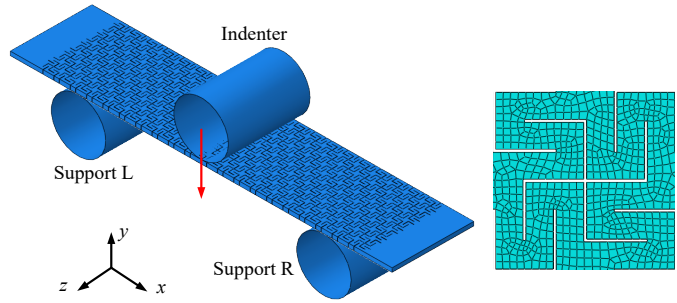

Fig. 4. Finite element model of the chiral lattice with unit cell a=6b under three-point bending test.

## 5. Results and discussion

### 5.1 Comparisons of the effective bending modulus $E_f$ amongst theoretical, finite element and experimental data

The results from the theoretical, numerical simulation and experimental tests related to the flexural modulus are shown in Table 1. The analytical bending modulus of the chiral fractal metamaterial with the configuration $a=6b$ differs from the numerical result by 6.6%; the related three-point bending experimental results are 4.5% higher and 2.0% lower than those from the theoretical and finite element predictions, respectively. The metamaterial with a unit cell $a=10b$ shows an experimentally derived flexural modulus that is 4.7% and 0.3% lower than the one provided by the analytical and numerical models. The discrepancy between the analytical and finite element model predictions for the chiral fractal lattice with unit cell $a=10b$ is 4.4%. The differences observed here have various explanations. The analytical model of the unit cell is based on an equivalent and continuously unfolded cantilever beam, while the full-scale finite element models and experimental samples are based on simply supported beams. The effect of using the Timoshenko beam with a shear coefficient in the analytical model also needs to be considered. The chiral fractal unit cell with $a=6b$ is described by beam elements, all regarded as Timoshenko beams. The transverse shear stiffness of those beams tends to increase the compliance under bending; as a result, the analytical model predicts an effective bending modulus lower than the one provided by the full-scale finite element model and the experimental tests. The finite element model provides only a slightly more accurate prediction compared to the experimental results, indicating that the analytical solution without the contact model is still valid for small out-of-plane bending deformations, and the contact friction on the interfaces does not have a significant influence on the deformation of the thin chiral lattice structure. When the presence of a Euler-Bernoulli beam was considered in the analytical model only, the resulting analytical bending moduli for the unit cells with $a=6b$ and $a=10b$ were 221.70MPa and 49.29 MPa, respectively.

**Table 1 Comparison between theoretical, finite element and experimental results of the effective bending modulus for the chiral lattice structure with unit cells a=6b and a=10b.**

| $E_f$ (MPa) | Analytical | FE | Experiment |
|---|---|---|---|
| a=6b | 215.62 | 229.78 | 225.25±3.89 |
| a=10b | 48.86 | 46.07 | 45.94±1.89 |

Force-displacement curves of the chiral fractal metamaterial with unit cell a=6b and a=10b from finite elements and experiments are displayed in Fig. 5 (a) and (b). The curves again demonstrate the good agreement between numerical and experimental results. It is worth noticing that the force-displacement curves do not show any hard stiffening nonlinearity, as might be expected if the contact between ribs occurred. This is likely due to the deformation being too small to cause contact. The experimental force-displacement curves of experiments show a clear elastic response followed a slight softening with the increase of the displacement. Force-displacement curves of the chiral fractal metamaterial with different unit cells and an integral plate, all made with the same PMMA substrate, are further compared within the same controlled displacement range (Fig. 5 (c)). As it will be further discussed in the following paragraphs, the bending stiffness of PMMA is 10 times and 51 times larger than that of chiral fractal metamaterial with unit cells a=6b and a=10b having the same dimensions, respectively.

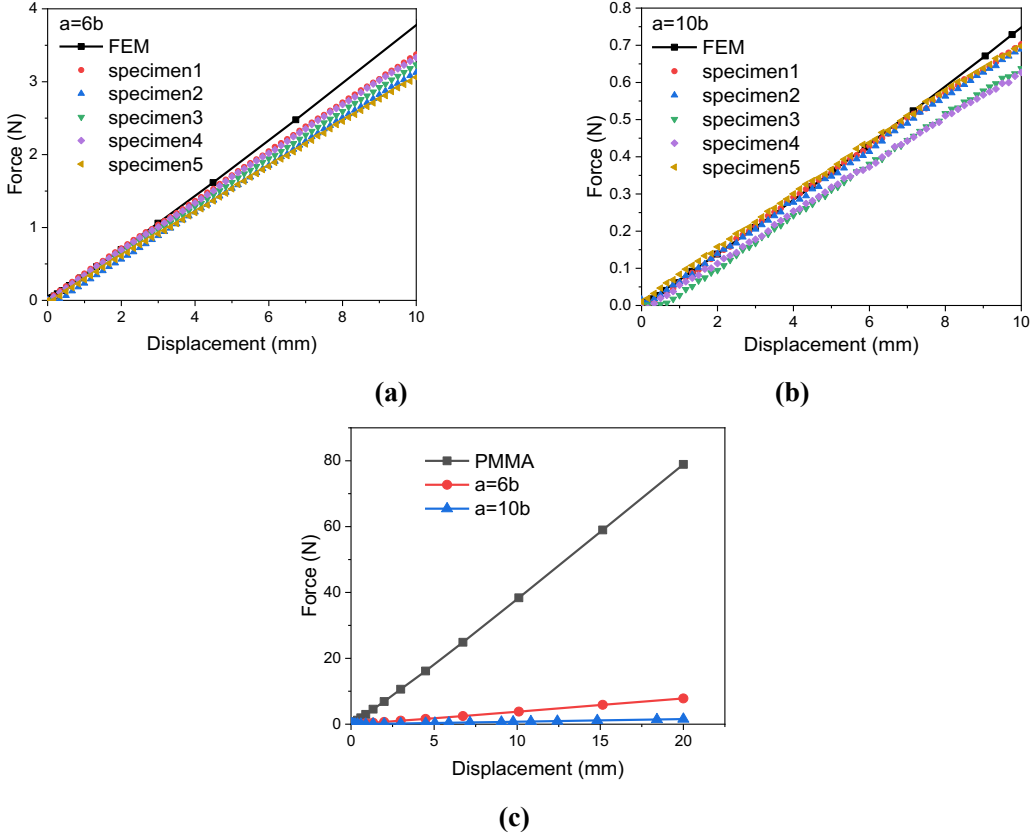

**Fig. 5 Force-displacement curves of chiral fractal metamaterial with unit cell a=6b (a) and a=10b (b), as well as the PMMA integral substrate with same dimensions (c).**

The deformations of the finite element models representing the chiral fractal mechanical metamaterial beam with unit cell of *a=6b* under three-point bending are shown in Fig. 6. The maximum von Mises stress of the metamaterial is only 17.5 MPa when the central displacement $U_2$ reaches 20mm; this indicates that the rotationally symmetric patterns of the slits in the unit cells of the metamaterial extend the overall linear elasticity stage of the PMMA substrate of the chiral structure (Fig. 6 (a)). The experimental test did not reach failure of the metamaterial beam. Fig. 6 (b) to Fig. 6 (g) show the contours of all the stress components along the three directions. The $\sigma_{11}$ stress along the *x* direction is at least twice larger than the other stresses along the *y* and *z* directions, which makes that stress the main contributor to the bending deformation (Fig. 6 (b)-(d)). Similarly, the shear stress $\sigma_{13}$ in *xz* plane is 1.5 times higher than the other shear stresses in the *yz* and *xy* planes, respectively (Fig. 6 (e)-(g)). The maximum shear force $\sigma_{13}$ in one unit cell occurs on the diagonal ends of the beam elements and along 45 degree (Fig. 6 (g)).

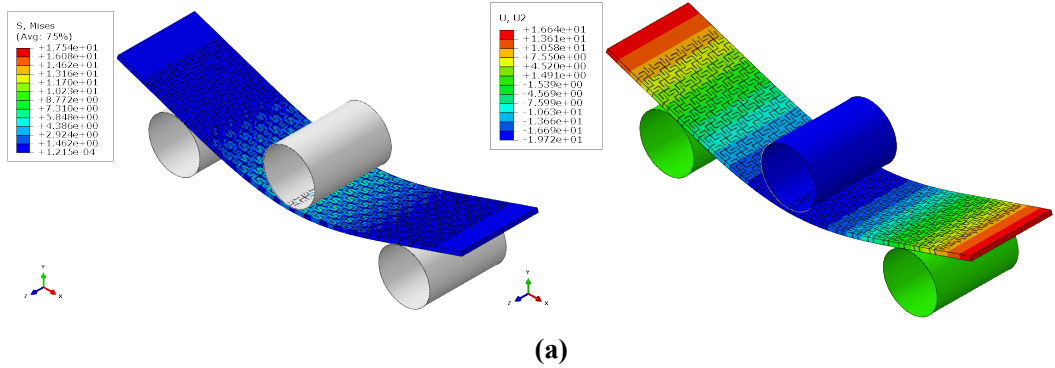

**(a)**

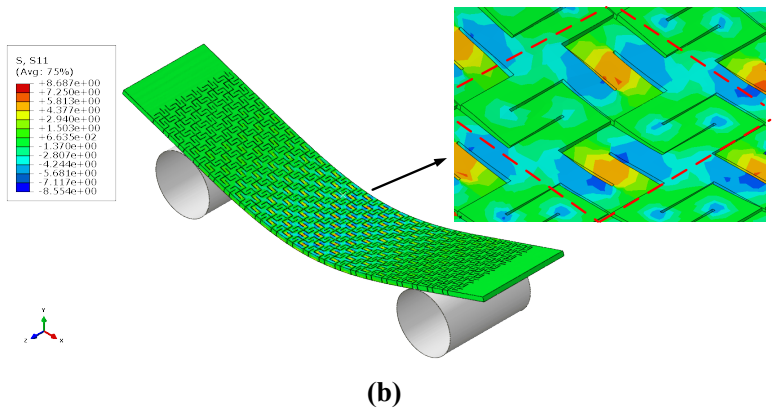

**(b)**

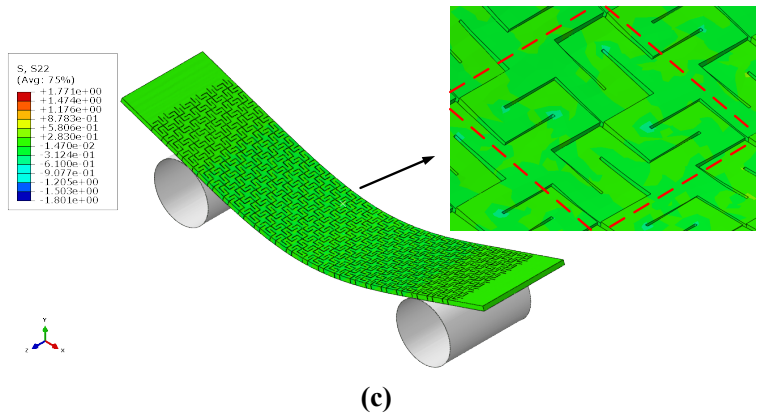

**(c)**

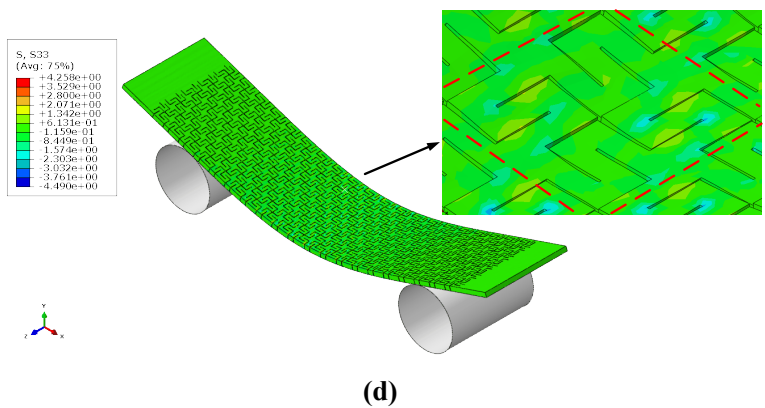

**(d)**

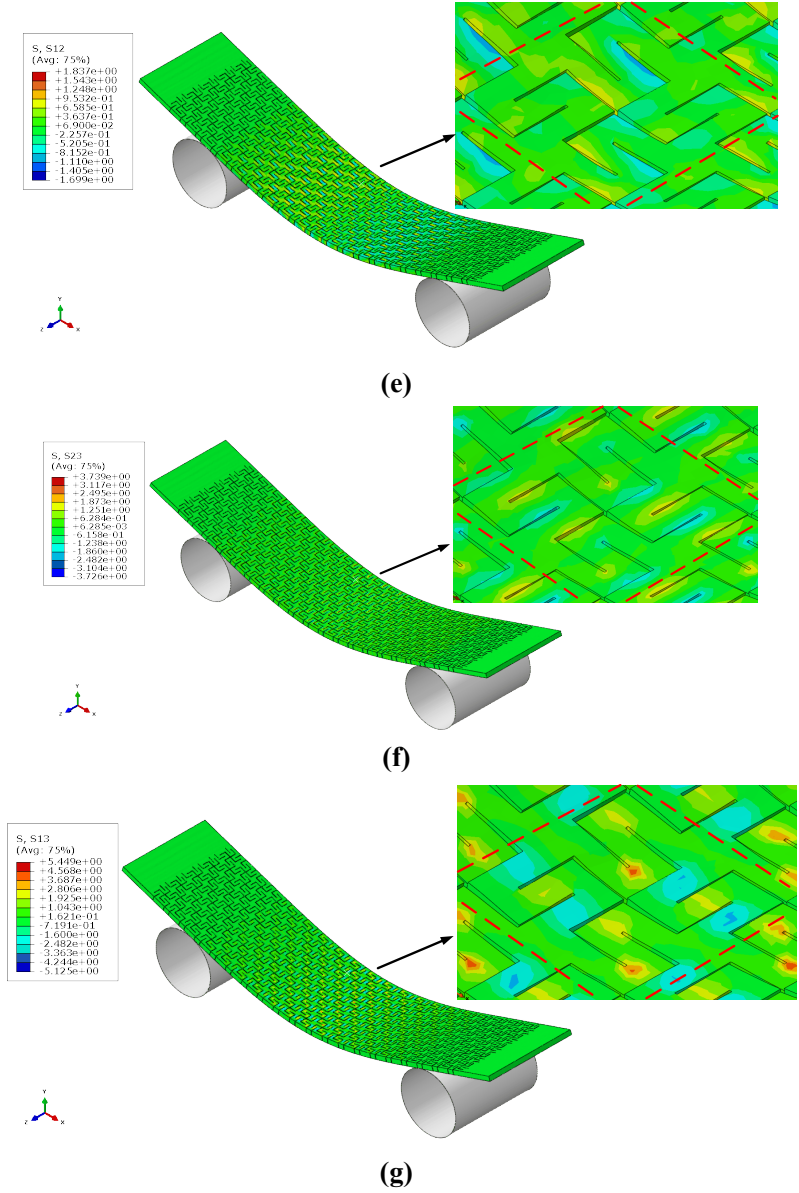

**Fig. 6.** Numerical results of chiral fractal metamaterial beams with unit cell a=6b under three-point bending under a central displacement U$_2$ of 20mm: (a) von Mises stress, (b) σ$_{11}$, (c) σ$_{22}$, (d) σ$_{33}$, (e) σ$_{12}$, (f) σ$_{23}$, (g) σ$_{13}$ (Stresses are in MPa).

### 5.2 Effect of the aspect ratio *a/b*

The magnitude of the in-plane engineering constants of the chiral fractal lattice metamaterial is strongly dependent upon the aspect ratio *a/b* between internal slits and ribs (Fig. 7) [39]. This is also true for the flexural modulus, as confirmed by the parametric analysis from the analytical model and the finite element simulations. We first look at the non-dimensional effective bending modulus $E_f/E_c$ (with $E_c$ being again the modulus of the solid substrate) versus the ratio between the unit cell width to out-plane thickness *b/h*, for the different slit to ribs aspect ratios *a/b* (Fig. 8). These parametric results are extracted from the analytical model. For a constant value of $b/h$, the nondimensional flexural modulus $E_f/E_c$ decreases rapidly with the increasing numbers of slits $a/b$. Fig. 8 (a) shows that when $b/h$ is less than 1, $E_f/E_c$ increases drastically for small values of $a/b$. The curves in Fig. 8 (a) shown in solid lines represent the cases for which a minimum slenderness ratio of 5 is considered, to make the equivalent beam of the unit cell satisfying the minimum requirements to be a Timoshenko beam. This condition is imposed by the relation $a/b \times b/h = 5$. When $b/h=1$, the non-dimensional bending modulus for the unit cell with a=6b is almost 5 times larger than the one of the unit cells with a=14b. However, the ratio $b/h$ has almost no effect on $E_f/E_c$ for different values of $a/b$ when $b/h > 5$ (Fig. 8 (b)). The nondimensional flexural modulus the chiral fractal lattice metamaterial with $a=6b$ varies between 0.156 and 0.161, while it is between $2.39 \times 10^{-2}$ and $2.46 \times 10^{-2}$ when the unit cell has $a=14b$. The variation of $E_f/E_c$ between unit cell with $a=10b$ and $a=14b$ is lower, compared to those existing between unit cells having $a=10b$ and $a=6b$.

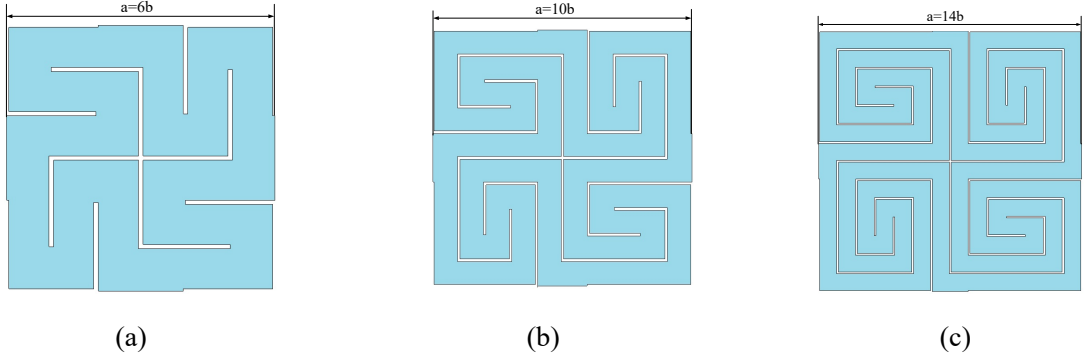

(a)          (b)          (c)

**Fig. 7. Three types of representative unit cells (RUC) with different aspect ratios of slit to ribs: (a) *a/b*=6, (b) *a/b*=10, (c) *a/b*=14**

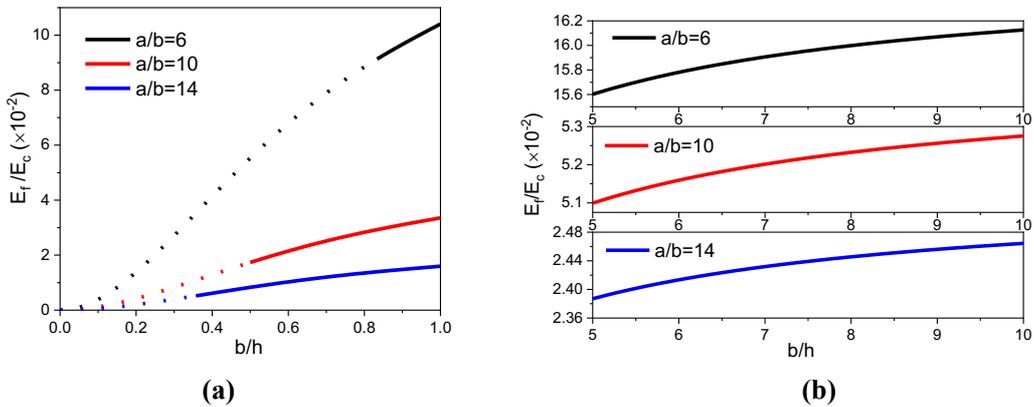

(a)          (b)

**Fig. 8.** Theoretical non-dimensional effective flexural modulus $E_f/E_c$ versus the ratio between width and out-plane thickness *b/h* for different aspect ratios of slit to ribs: (a) $1 \geq b/h > 0$; (b) $b/h > 5$

The effects on the effective bending modulus provided by the torque, bending moment and shearing force present in the ribs of the chiral fractal metamaterials have been also investigated for different aspect ratios *a/b*. The comparison between the theoretical and the full-scale finite element results are shown in Fig. 9. In the analytical results, we differentiate the contribution from the torque (T), bending (B) and shear (S) induced deformations in the ribs. All the analytical models - except the one including the combined bending and shear loading (BS) - provide consistent results with the high-fidelity numerical simulations for the different aspect ratios *a/b*. The shear deformation within the ribs contributes significantly to the equivalent bending modulus, especially for metamaterials with unit cells having the lowest aspect ratio *a/b*=6. The bending modulus increases 3.4 times compared to the finite element case, when the torque-induced deformation is neglected. Similar to the in-plane mechanical properties [39], high aspect ratios *a/b* lead to a decrease in bending stiffness. Quite interestingly, the relation of $E_f/E_c$ versus *a/b* in the log-log scale shows a good linearity, especially for the highest fidelity models. Linearity in log-log scales of fractal entities versus systems parameters has been already observed, for example, in diffusion coefficients associated with stochastic ensembles of moving particles [45], and in disordered-averaged entropy systems with fractal entanglement scales [46].

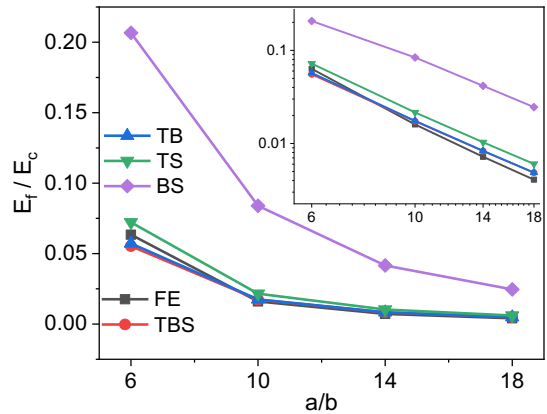

**Fig. 9.** Comparisons between theoretical and finite element results related to the nondimensional flexural modulus *$E_f$ / $E_c$* for different slit/ribs length ratios of chiral fractal metamaterials. The theoretical models consider different micromechanisms: TB stands for torque/bending, TS for torque/shear, BS for bending/shear and TBS for **torque/bending/shear. The graph in the inset shows the nondimensional flexural modulus versus *a/b* in logarithmic scale.**

It is notable to observe the variation of the ratio between the out-plane effective bending modulus $E_f$ to the in-plane effective Young's modulus $E_x$ [39] for different aspect ratios *a/b* (Fig. 10 (a)). The results from the theoretical models combine all the microstructure mechanisms considered acting within the unit cell of the chiral fractal metamaterials (BAS: in-plane Bending moment, Axial and in-plane Shear force under in-plane tension; TBS: Torque, out-of-plane bending Moment and Shear force for out-plane bending). All the models also agree well with finite element data representing full-scale tensile and bending tests. The torque and out-of-plane bending moment, as well as the axial and in-plane bending moment, all provide the most important contributions to the ratio between effective bending and tensile moduli of the chiral fractal metamaterial with different slit/ribs length ratios *a/b*. The ratios between moduli vary from ~ 5 for *a/b*=6, to ~ 34 for *a/b*=18. Architectures like double-twill carbon/flax fabric and epoxy laminates have shown ratios between flexural and tensile moduli up to ~ 4.1 [47]. Cactus fibres, which have a fractal and tree-like configuration, have ratios between equivalent bending and tensile moduli up to ~ 6.7 [48]. These values are large and allow tuning by variation of the fractal geometry: changing the aspect ratio changes the number of loops in a unit cell. By contrast, the difference between bending and tensile stiffness in hexagonal honeycomb [30], [31] depends only on the thickness *t* of the cell walls in relation to their length *L*, as expressed by the ratio *t/L*. Bending stiffness is proportional to *t/L* by contrast to the well-known axial in plane stiffness proportional to $(t/L)^3$. The analogous for the chiral fractal metamaterials of the ratio *t/L* is $(a/b)^{-1}$. The chiral fractal metamaterials show indeed a linear proportionality between the bending to axial moduli ratio and $(a/b)^2$, as it can be evinced from the $R^2 = 0.99$ fitting of the numerical results (Fig. 10 (b)). This is like the $(t/L)^{-2}$ linear dependence of the flexural to tensile stiffness in lattices and honeycombs. Large effects of the order of 30 were observed in lattices, even non chiral ones [49]. Also, designed lattices exhibit strong size effects in bending and torsion [50]; slender specimens in which gradients are large are a factor 29 to 36 stiffer than in the absence of gradients, corresponding to pure tension or pure shear. The chiral fractal metamaterial described here shows however that large bending to axial tensile moduli ratios can be achieved by purely tuning the fractal order of the architecture.

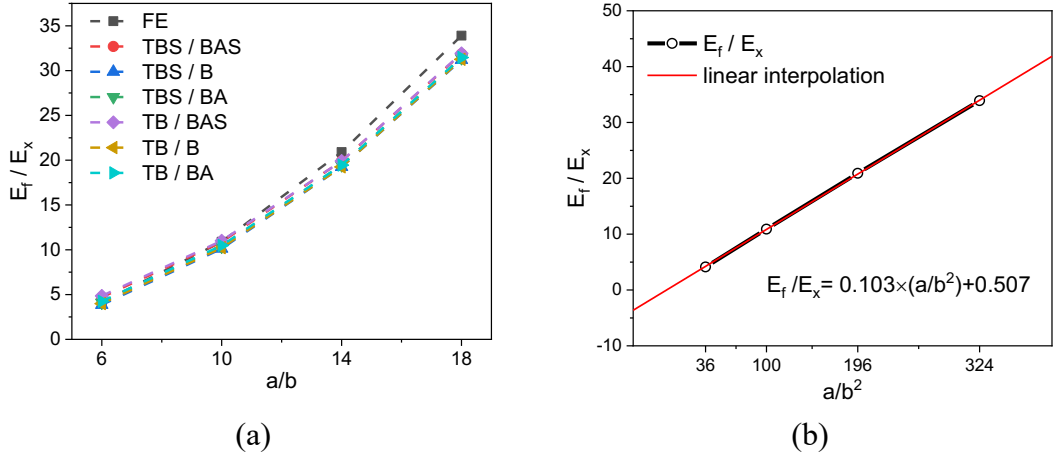

(a)                                (b)

**Fig. 10.** (a) Ratio between effective bending and tensile (Young's) moduli $E_f / E_x$ versus the different slit/rib length ratios of the chiral fractal metamaterials. (b) Linear dependence of the ratio between bending and tensile stiffness (finite element data), and the square of $a/b$.

## 6. Conclusions

This work has focused on the out-plane bending of chiral fractal lattice perforated metamaterials with the use of theoretical and finite element models, together with experimental tests. Like its in-plane tensile modulus, the flexural modulus of these chiral fractal metamaterials decreases with the increase of the fractal parameter of the cells $a/b$, with an almost linear dependence in terms of log-log scales, similarly to other fractal systems shown in open literature. The parametric analysis performed shows that, within the limit of the use of Euler-Bernoulli and/or Timoshenko beams to represent the components of the internal architecture of the chiral fractal units, the overall bending modulus is almost independent from the transverse thickness of the cell. Large bending to axial moduli ratios that depend on the fractal order are observed; these metamaterials show therefore the possibility of being used as platforms to design architected structures with differential bending to axial stiffness properties, by tailoring the topology of the metamaterial using a suitable fractal order.

### Acknowledgements

WZ and DZ would like to thank Chinese Scholarship Council (CSC) for the funding of their research work through the University of Bristol. RN also acknowledges the support from UK Engineering and Physical Sciences Research Council (EPSRC

EP/N509619/1) for his Postdoctoral Fellowship. JY would like to acknowledge the support of the Small Research Grant from Royal Society of Edinburgh (RSE/1754). FS would like to thank the European Commission for the computational and testing logistics provided through the H2020-1.3.1.675441 MSC ITN VIPER and ERC-AdG-H2020 101020715 NEUROMETA projects.

**References**

[1] Gibson, L J and Ashby, M F. Cellular Solids: Structure and Properties. 2nd Edition, Cambridge University Press, Cambridge, 1997.

[2] Grima JN, Evans KE (2000) Auxetic behavior from rotating squares. J Mater Sci Lett 19:1563–1565.

[3] Grima JN, Alderson A, Evans KE (2004) Negative Poisson's ratios from rotating rectangles. Comp Methods Sci Technol 10(2):137–145.

[4] JT. Overvelde, S. Shan, K. Bertoldi. Compaction through buckling in 2D periodic, soft and porous structures: effect of pore shape. Advanced Materials. 2012, 24(17), 2337-2342.

[5] M. Taylor, L. Francesconi, M. Gerendas, A. Shanian, C. Carson, K. Bertoldi. Low porosity metallic periodic structures with negative Poisson's ratio. Advanced Materials. 2014, 26(15), 2365-70.

[6] J. Shen, S. Zhou, X. Huang, Y. M. Xie. Simple cubic three-dimensional auxetic metamaterials. Physica Status Solidi, 2014, 251(8):1515-1522.

[7] B. Florijn, C. Coulais, M. van Hecke. Programmable mechanical metamaterials. Physical Review Letters, 2014, 113(17):175503.

[8] R. Gatt, L. Mizzi, J. I. Azzopardi, K. M. Azzopardi, D. Attard, A. Casha, J. Briffa, J. N. Grima. Hierarchical Auxetic Mechanical Metamaterials. Scientific Reports, 2015, 5, 8395.

[9] SH Kang, S Shan, A Košmrlj, WL Noorduin, S Shian. Complex ordered patterns in mechanical instability induced geometrically frustrated triangular cellular structures. Phys Rev Lett, 2014, 112(9): 098701.

[10] A. Slann, W. White, F. Scarpa, K. Boba, I. Farrow. Cellular plates with auxetic rectangular perforations. Physica Status Solidi, 2015, 252(7):1533-1539.

[11] S Shan, SH Kang, P Wang, C Qu, S Shian. Harnessing multiple folding mechanisms in soft periodic structures for tunable control of elastic waves. Advanced Functional Materials, 2015, 24(31):4935-4942.

[12] Ahmad Rafsanjani, Damiano Pasini. Bistable auxetic mechanical metamaterials inspired by ancient geometric motifs. Extreme Mechanics Letters 9 (2016) 291–296.

[13] Kévin Billon, Ioannis Zampetakis, Fabrizio Scarpa, Morvan Ouisse, Emeline Sadoulet-Reboul, Manuel Collet, Adam Perriman, Alistair Hetherington. Mechanics and band gaps in hierarchical auxetic rectangular perforated composite metamaterials. Composite Structures, 2017, 160, 1042–1050.

[14] J. Shim, S. Shan, A. Kosmrlj, S. H. Kang, E. R. Chen, J. C. Weaver, K. Bertoldi. Harnessing instabilities for design of soft reconfigurable auxetic/chiral materials. Soft Matter 2013, 9(34), 8198-8202.

[15] L Mizzi, K.M. Azzopardi, D Attard, JN Grima, R Gatt. Auxetic metamaterials exhibiting giant negative Poisson's ratios. Physica status solidi (RRL)-Rapid Research Letters. 2015, 9(7):425-430

[16] Cho Y, Shin J-H, Costa A, Kim TA, Kunin V, Li J, et al. Engineering the shape and structure of materials by fractal cut. Proc Natl Acad Sci 2014;111 (49):17390–5.

[17] S. Shan, S. H. Kang, Z. Zhao, L. Fang, K. Bertoldi. Design of planar isotropic negative Poisson's ratio structures. Extreme Mechanics Letters, 2015, 4:96-102.

[18] J. N. Grima, L Mizzi, K.M. Azzopardi, R Gatt. Auxetic Perforated Mechanical Metamaterials with Randomly Oriented Cuts. Advanced Materials, 2016, 28, 385–389

[19] Y Tang, G Lin, L Han, S Qiu, S Yang. Design of Hierarchically Cut Hinges for Highly Stretchable and Reconfigurable Metamaterials with Enhanced Strength. Advanced Materials, 2015, 27(44):7181-90.

[20] H Mohanraj, SLM Filho Ribeiro, TH Panzera, F Scarpa, IR Farrow. Hybrid auxetic foam and perforated plate composites for human body support. Physica Status Solidi, 2016, 253(7):1378-1386.

[21] Zhennan Zhang, Fabrizio Scarpa, Brett A. Bednarcyk, Yanyu Chen. Harnessing fractal cuts to design robust lattice metamaterials for energy dissipation. Additive Manufacturing 46 (2021) 102126.

[22] R. LAKES. Deformation mechanisms in negative Poisson's ratio materials: structural aspects. Journal of Materials Science 26 (1991) 2287-2292.

[23] D.Prall, R.S.Lakes. Properties of a chiral honeycomb with a Poisson's ratio of -1. Int. J. Mech. Sci. 1997, 39(3), 305-314.

[24] H. Abramovitch, M. Burgard, Lucy Edery-Azulay, K.E. Evans, M. Hoffmeister, W. Miller, F. Scarpa, C.W. Smith, K.F. Tee. Smart tetrachiral and hexachiral honeycomb: Sensing and impact detection. Composites Science & Technology, 2010, 70(7):1072-1079.

[25] Miller W, Smith CW, Scarpa F, Evans KE. Flatwise buckling optimization of hexachiral and tetrachiral honeycombs. Compos Sci Technol 2010;70 (7):1049–56.

[26] Spadoni A, Ruzzene M, Scarpa F. Global and local linear buckling behaviour of a chiral cellular structure. Phys Status Solidi B 2005;242(3):695–709.

[27] Scarpa F, Blain S, Lew T, Perrott D, Ruzzene M, Yates JR. Elastic buckling of hexagonal chiral cell honeycombs. Composites Part A 2007;38(2):280–9.

[28] A. Lorato, P. Innocenti, F. Scarpa, A. Alderson, K.L. Alderson, K.M. Zied, N. Ravirala, W.Miller, C.W. Smith , K.E. Evans. The transverse elastic properties of chiral honeycombs. Composites Science and Technology 70 (2010) 1057–1063.

[29] Alderson A, Alderson KL, Chirima G, Ravirala N, Zied KM. The in-plane linear elastic constants and out-of-plane bending of 3-coordinated ligament and cylinder-ligament honeycombs. Compos Sci Technol 2010;70(7):1034–41.

[30] D.H. Chen. Bending deformation of honeycomb consisting of regular hexagonal cells. Composite Structures 93 (2011) 736–746.

[31] D.H. Chen. Equivalent flexural and torsional rigidity of hexagonal honeycomb. Composite Structures 93 (2011) 1910–1917.

[32] Y. Hou, Y.H.Tai, C.Lira, F.Scarpa, J.R.Yates, B.Gu. The bending and failure of sandwich structures with auxetic gradient cellular cores. Composites: Part A49(2013)119–131.

[33] C. S. Ha, M.E. Plesha, R.S. Lakes. Chiral three-dimensional lattices with tunable Poisson's ratio. Smart Materials and Structures, 2016, 25(5):054005.

[34] Lakes, R.S, Benedict, R.L. Noncentrosymmetry in micropolar elasticity. International Journal of Engineering Science, 20 (10), 1161-1167, (1982).

[35] Jian Huang, Qiuhua Zhang, Fabrizio Scarpa, Yanju Liu, Jinsong Leng. Bending and benchmark of zero Poisson's ratio cellular structures. Composite Structures 152 (2016) 729–736.

[36] Dejan Mitov, Bojan Tepavčević, Vesna Stojaković, Ivana Bajšanski. Kerf Bending Strategy for Thick Planar Sheet Materials. Nexus Network Journal, 2019, 21, 149–160.

[37] Saeid Zarrinmehr, Mahmood Ettehad, Negar Kalantar, Alireza Borhani, Shinjiro Sueda, Ergun Akleman. Interlocked archimedean spirals for conversion of planar rigid panels into locally flexible panels with stiffness control. Computers & Graphics, 2017, 66, 93-102.

[38] Caleb Widstrand, Negar Kalantar, Stefano Gonella. Bandgap tuning in kerfed metastrips under extreme deformation. Extreme Mechanics Letters 53 (2022) 101693.

[39] W.Zhang, Robin Neville, Dayi Zhang, Fabrizio Scarpa, Lifeng Wang, Roderic Lakes. The two-dimensional elasticity of a chiral hinge lattice metamaterial. International Journal of Solids and Structures, 2018, 141–142, 254–263.

[40] G Peano. Sur une courbe, qui remplit toute une aire plane. Math. Ann., 36 (1) (1890), 157-160.

[41] Young WC, Budynas RG. Roark's formulas for stress and strain. New York: McGraw-Hill; 2002.

[42] Cowper, GR, 1966. The shear coefficient in Timoshenko's beam theory. J. Appl. Mech. 33 (2), 335e40.

[43] M Najafi，H Ahmadi，GH Liaghat. Investigation on the flexural properties of sandwich beams with auxetic core. Journal of the Brazilian Society of Mechanical Sciences and Engineering (2022) 44:61.

[44] L Meng，X Lan，J Zhao，H Li，L Gao. Failure analysis of bio-inspired corrugated sandwich structures fabricated by laser powder bed fusion under three-point bending. Composite Structures 263 (2021) 113724.

[45] N Korabel，R Klages，AV Chechkin，IM Sokolov，VY Gonchar. Fractal Properties of Anomalous Diffusion in Intermittent Maps. Phys. Rev. E, 2007, 75, 036213.

[46] M Ippoliti，T Rakovszky，V Khemani. Fractal, Logarithmic, and Volume-Law Entangled Nonthermal Steady States via Spacetime Duality. Phys. Rev. X, 2022, 12, 011045.

[47] V Fiore, A Valenza, G Di Bella. Mechanical behavior of carbon/flax hybrid composites for structural applications. Journal of Composite Materials, 2011, 46(17), 2089–2096.

[48] M Bouakba, A Bezazi, K Boba, F Scarpa, S Bellamy. Cactus fibre/polyester biocomposites: Manufacturing, quasi-static mechanical and fatigue characterisation. Composites Science and Technology 74 (2013) 150–159.

[49] Rueger, Z, Lakes, R. S. Strong Cosserat elasticity in a transversely isotropic polymer lattice. Physical Review Letters, 120, 065501 Feb. (2018).

[50] Rueger, Z, Lakes, R. S. Strong Cosserat elastic effects in a unidirectional composite. Zeitschrift für angewandte Mathematik und Physik 68, 54 (2017).